\documentclass[twocolumn,showpacs,preprintnumbers,amsmath,amssymb,nofootinbib,floatfix]{revtex4}

\usepackage[dvips]{graphicx}
\usepackage{dcolumn}
\usepackage{bm}

\makeatletter
 
  \@addtoreset{equation}{section}
\makeatother

\begin{document}

 \title{Excited Gauge and Higgs Bosons in the Unified Composite Model}

\author{Hidezumi Terazawa}
\email{terazawa@mrj.biglobe.ne.jp}
\affiliation{\vspace{5mm}%
\sl Center of Asia and Oceania for Science(CAOS), 3-11-26 Maesawa, Higashi-kurume, Tokyo 203-0032, Japan\\
and\\
Midlands Academy of Business \& Technology(MABT), Mansion House, 41 Guildhall Lane, Leicester LE1 5FR, United Kingdom
}

\author{Masaki Yasu\`{e}}%
\email{yasue@keyaki.cc.u-tokai.ac.jp}
\affiliation{\vspace{5mm}%
\sl Department of Physics, Tokai University, 4-1-1 Kitakaname,\\
Hiratsuka, Kanagawa 259-1292, Japan
}


\begin{abstract}
In the unified subquark model of all fundamental particles and forces, the mass of the Higgs boson in the standard model 
of electroweak interactions ($m_H$) is predicted to be about $2\sqrt{6}m_W/3$ (where $m_W$ is the mass of the charged 
weak boson, $W$), which agrees well with the experimental values of about 125 GeV recently found by the ATLAS and CMS 
Collaborations at the LHC. It seems to indicate that the Higgs boson is a composite of the iso-doublet spinor subquark-antisubquark 
pairs well described by the unified subquark model with either one of subquark masses vanishing or being very small 
compared to the other. In the unified composite model, the masses of excited weak gauge bosons, $W^\ast$ and $Z^\ast$, and Higgs 
bosons, $H^\ast$, as well as excited quarks and leptons are all predicted to be of the order of the composite mass scales, 
say 1 TeV, and to satisfy the relation of $m_Wm_{W^\ast} = m_Zm_{Z^\ast}\cos\theta_w$ (where $\theta_w$ is the weak mixing angle). 
It strongly suggests that the excess of di-boson resonance events recently found by the ATLAS Collaboration at about 2 TeV 
may be explained by the productions and decays of the first excited states of either one of the weak and Higgs bosons, and 
the glueballs. 
\end{abstract}
\maketitle
\section{\label{sec:1}Introduction}
What most of us could expect to find in high energy experiments at the Large Hadron Collider was the Higgs boson ($H$), 
which was the only fundamental particle that had not been found in the standard model of electroweak interactions \cite{STD}. 
In the unified subquark model of all fundamental particles and forces \cite{Subquarks}, the mass of the Higgs boson has been predicted 
in the following way: 

In general, in composite models of the Nambu-Jona-Lasinio type \cite{NJL}, the Higgs boson appears as a composite state of fermion-antifermion 
pairs with the mass twice as much as the fermion mass. The unified subquark model of the Nambu-Jona-Lasinio type
\cite{NJLSubquark} has predicted the following two sum rules:
\begin{equation}
m_W=[3(m_{w_1}^2+m_{w_2}^2)/2]^{1/2},
\label{mW_Sumrule1}
\end{equation}
and
\begin{equation}
m_H=2[(m_{w_1}^4+m_{w_2}^4)/(m_{w_1}^2+m_{w_2}^2)]^{1/2}, 
\label{mH_Ssumrule}
\end{equation}
where $m_{w_1}$ and $m_{w_2}$ are the masses of the iso-doublet spinor subquarks called \lq\lq wakems\rq\rq\ standing 
for weak and electromagnetic ($w_i$ for $i=1,2$) while $m_W$ and $m_H$ are the masses of the charged weak boson ($W$) and 
physical Higgs boson in the standard model, respectively. By combining these sum rules, the following relation has been 
obtained if $m_{w_1}=m_{w_2}$:
\begin{equation}
m_w:m_W:m_H=1:\sqrt{3}:2.
\label{mw:mW:mH_1}
\end{equation}
From this relation, the wakem and Higgs boson masses have been predicted as
\begin{equation}
m_w=m_W/\sqrt{3}=46.4 ~{\rm GeV},
\label{Predicted_mw_1}
\end{equation}
and
\begin{equation}
m_H=2m_W/\sqrt{3}=92.8 ~{\rm GeV},
\label{Predicted_mH_1}
\end{equation}
for $m_W=80.4$ GeV \cite{PDG}.
On the other hand, if $m_{w_1}=0$ or $m_{w_2}=0$, the other relation can be obtained:
\begin{equation}
m_w:m_W:m_H=1:\sqrt{3/2}:2.
\label{mw:mW:mH_2}
\end{equation}
From this relation, the non-vanishing wakem and Higgs boson masses can be predicted
\begin{equation}
m_w=m_W/\sqrt{3/2}=65.6 ~{\rm GeV},
\label{Predicted_mw_2}
\end{equation}
and
\begin{equation}
m_H=2m_W/\sqrt{3/2}=131 ~{\rm GeV},
\label{Predicted_mH_2}
\end{equation}
for $m_W=80.4$ GeV \cite{PDG}.
More generally, from the two sum rules, the Higgs boson mass can be bounded as 
\begin{equation}
92.8~{\rm GeV}=2m_W/\sqrt{3}\leq m_H\leq 2\sqrt{6}m_W/3=131~{\rm GeV}.
\label{mH_Bounded}
\end{equation}

Recently, the ATLAS and CMS Collaboration experiments at the CERN Large Hadron Collider have found the Higgs boson 
with the mass at about 125 GeV \cite{HiggsATLAS,HiggsCMS}, very close to the predicted one of $m_H = 2\sqrt{6}m_W/3 = 131$ GeV. 
It seems to indicate that the Higgs boson is a composite of the iso-doublet spinor subquark-antisubquark pairs 
well described by the unified subquark model with either one of subquark masses vanishing or being very small compared 
to the other \cite{HiggsSubquark}. 

What is the energy scale of compositeness for the composite Higgs boson? If the Higgs boson is a composite state of 
subquark-antisubquark pairs bound by the confining force due to the Yang-Mills gauge fields of subcolor $SU(N)$ \cite{Subcolor}, 
the composite energy scale, $\Lambda_{sc}$, is analogous to $\Lambda_c$ (which is of the order about 100 MeV) and is rather 
arbitrary. If it is a composite state of techniquark-anti-techniquark pairs, the composite energy scale is given either by  
$\Lambda_{TC}$ (where $TC$ stands for technicolor) or by $m_{tq}$ (where $tq$ stands for techniquark), which is 
of the order of 1 TeV  \cite{Technicolor}. If it is composite state of topquark-anti-topquark pairs \cite{Topquark}, the composite energy scale is 
given by the topquark mass, $m_t$, which is about $174$ GeV \cite{PDG}.

Once the energy scale of compositeness, $\Lambda$, is given, it seems natural to expect that there exist not only 
the excited Higgs boson but also the excited weak bosons as well as excited quarks and leptons all at the mass scale of 
$\Lambda$, say 1 TeV. This is true in the unified composite model and a consequence of what is called the principle 
of triplicity for hadrons, quarks, and subquarks \cite{Triplicity}. The energy scale of compositeness of the order of 1 TeV is 
also crucial in explaining the mass hierarchy of the first, second and third generations of quarks and leptons \cite{Generation}. 
In 1982, we already discussed various possible effects of the substructure of not only quarks and leptons 
but also gauge and Higgs bosons including excited gauge and Higgs scalars \cite{SubquarkPhenomena}.

The purpose of this paper is to update the theoretical expectations on these excited states of not only gauge and Higgs 
bosons but also glueballs for future experimental studies. 

\section{\label{sec:2}Excited Weak Bosons}
Excited weak bosons \cite{StrongWeakBoson}, $W^\ast$ and $Z^\ast$, are the excited states of the weak bosons, $W$ and $Z$, in the $SU(2)_L\times U(1)_Y$ model \cite{STD} that can decay into $W\gamma$ and $Z\gamma$, respectively.  In the unified subquark model \cite{Subquarks,NJLSubquark}, $W$ and $Z$ originate from the weak-isospin triplet gauge bosons of $SU(2)_L$, $V^i$ ($i$=1,2,3), consisting of wakem and antiwakem pairs.  If composite weak bosons include composite states consisting of iso-doublet scalar wakem ($\omega$) and antiwakem pairs, the created weak bosons are guaranteed to be massless composites since they behave as the gauge fields of $SU(2)_L$ \cite{ExcitedBosons,ExtraSU2}. Namely, $V^i$ described by
\begin{equation}
V^i_\mu\sim 
{\overline w_L} 
\gamma_\mu
\tau^i
w_L
+
\omega^\dagger 
\mathord{\buildrel{\lower3pt\hbox{$\scriptscriptstyle\leftrightarrow$}} 
\over \partial_\mu }
\tau^i
\omega,
\label{CompositeGaugeBoson}
\end{equation}
where $\tau^i$ stands for the Pauli matrix of $SU(2)_L$, exhibits the correct transformation of the gauge field induced by the $SU(2)_L$ gauge transformation of $w_L$ and $\omega$.  Similarly for the $U(1)_Y$ composite gauge field, $B_\mu$.  If this is the case, there is another composite state, $V^{i \ast}$, consisting of $w_L$:
\begin{equation}
V^{i\ast}_\mu\sim 
{\overline w_L} 
\gamma_\mu
\tau^i
w_L,
\label{CompositeBoson}
\end{equation}
which can be sources of $W^\ast$ and $Z^\ast$.

The effect of such massive states of $W^\ast$ and $Z^\ast$ can be estimated by the effective lagrangian approach \cite{EffectiveLagrangian}.  The contribution of the subquarks to the quark-lepton physics is known to be well evaluated by the effective lagrangian approach based on the Nambu-Jona-Lasinio type model \cite{NJLSubquark}.  It has also been discussed that the $SU(2)$ subcolor confining phase is equivalent to the Higgs phase, where subquark interactions are nothing but the Nambu-Jona-Lasinio type supplemented by auxiliary fields \cite{ExtraSU2,Equivalence1}. This equivalence is an explicit example of the \textit{complementarity} \cite{Subcolor,Complementarity} as well as the \textit{transmuted gauge symmetry} \cite{Transmuted}. Using $w$ and $\omega$ as the relevant subquarks, one can find the effective interactions of $V$ and $V^\ast$, which contain the kinetic mixing between the field strengths of $V$ and $V^\ast$. After this mixing is removed, $V$ and $V^\ast$ possessing diagonal kinetic terms are, respectively, denoted by ${\mathcal V}$ and ${\mathcal V}^\ast$, which, together with $B$, finally become $W$, $Z$, $\gamma$, $W^\ast$ and $Z^\ast$ \cite{ExcitedBosons}.  As a result, the effective interactions inducing $W^\ast\rightarrow W\gamma$ and $Z^\ast\rightarrow Z\gamma$ vanish to the leading order.  The next-to-leading order interactions arise from the effective interactions involving the $SU(2)_L$-doublet composite Higgs boson $\phi$ such as
\begin{equation}
\left({\mathcal D}_\mu \phi\right)^\dagger B^{\mu\nu}\left({\mathcal D}_\nu \phi\right)
\label{W+gamma1}
\end{equation}
with
\begin{equation}
{\mathcal D}_\mu = \partial_\mu+i\left( g{\mathcal V}_\mu+g^\ast\sqrt{1-\frac{g^2}{g^{\ast 2}}}{\mathcal V}^\ast_\mu\right)+ig^\prime\frac{1}{2}B_\mu,
\label{CovariantDerivative}
\end{equation}
where ${\mathcal V}_\mu= \tau^i{\mathcal V}_\mu^i/2$ and ${\mathcal V}_\mu^\ast= \tau^i{\mathcal V}_\mu^{i\ast}/2$, $B^{\mu\nu}$ is the field strength of $B_\mu$, $g$ ($g^\prime$) is the gauge coupling constant of the $SU(2)_L$ ($U(1)_Y)$ gauge field and $g^\ast$ is the coupling constant of ${\mathcal V}^\ast_\mu$ corresponding to $g$ of ${\mathcal V}_\mu$.  The compositeness conditions on $V$ and $V^\ast$ provide $g^\ast=\sqrt{3/2}g$.  More detailed discussions on $W^\ast\rightarrow W\gamma$ and $Z^\ast\rightarrow Z\gamma$ as well as on $W^\ast\rightarrow WZ$ and $Z^\ast\rightarrow WW, ZZ$ can be found in Ref.\cite{ExcitedBosons}.  

There is the celebrated mass relation among the masses of $W$, $Z$, $W^\ast$ and $Z^\ast$ \cite{ChargeCommutation,Rederived}:
\begin{equation}
m_Wm_{W^\ast}=m_Zm_{Z^\ast}\cos\theta_w,
\label{MassRelation}
\end{equation}
where $\theta_w$ is the weak mixing angle observed in low-energy neutrino-induced neutral-current reactions \cite{ExcitedBosons,EffectiveLagrangian}.  The expected phenomenology of new particles at LHC can be found in the literatures \cite{RecentStudyNewResonance} including discussions on the heavy $SU(2)_L$-triplet vector bosons such as $W^\ast$ and $Z^\ast$.  Since $m_W \sim m_Z\cos\theta_w$ holds experimentally, one can expect that $m_{W^\ast} \sim m_{Z^\ast}$. In fact, the relation predicts
\begin{equation}
0.997\lesssim \frac{m_{W^\ast}}{ m_{Z^\ast}}\lesssim 1.003,
\label{ExcitedMassRatio}
\end{equation}
for $m_W=80.386\pm 0.015$ GeV, $m_Z=91.1876\pm 0.0021$ GeV and $\sin^2\theta_w=0.2236\pm 0.041$ from the measured ratio of neutrino-induced neutral-to-charged-current cross sections \cite{PDG}.  There should appear $W^\ast$ and $Z^\ast$ almost with the same mass, which may have brought out $W^\ast\rightarrow WZ$ and $Z^\ast\rightarrow WW, ZZ$ at LHC \cite{ATLASsearch,ATLASanomaly,CMSsearch}.  In any case, detailed analyses of the decay modes of $W\gamma$ and $Z\gamma$ will be crucial in identifying $W^\ast$ and $Z^\ast$.

\section{\label{sec:3}Excited Higgs Bosons}
In the unified composite model, not only the weak bosons, $W$ and $Z$, but also the Higgs boson, $H$, is a composite state 
of wakem-antiwakem pairs bound due to the same confining force. Therefore, it seems natural to expect that there exist 
excited Higgs bosons at the same energy scale of compositeness, $\Lambda$. An excited Higgs boson, $H^\ast$, should decay into 
$H\gamma$, $WW$, $ZZ$, \textit{etc.} so that it may be found as a resonance of di-boson at $\Lambda$.

\section{\label{sec:4}Excited Glue-Balls}
The glueballs in a generic sense are hadrons which consist of gluons only. The simplest one of glueballs is a color singlet 
state consisting of two gluons, which we call glueball in a narrow sense. The second simplest one is a color-singlet state 
consisting of three gluons, which is so-called odderon. The one of the present authors (H.T.) has proposed a color-ball, the 
color-singlet complex object consisting of an arbitrary number of gluons \cite{Color-Balled Nuclei}. In the unified subquark model of 
all fundamental particles and forces \cite{Subquarks,NJLSubquark}, not only the weak and Higgs bosons but also the gluons are taken as composite 
states of subquark-antisubquark pairs. Therefore, it seems natural to expect that there exist excited glueballs 
or color-balls at the composite energy scale of $\Lambda$. Since the gluon-fusion process is a dominant process 
for particle productions at LHC, one can expect that excited color-balls would be copiously produced at LHC once the energy
becomes of the order of $\Lambda$.

\section{\label{sec:5}Conclusions}
In the unified composite model, the masses of excited weak and Higgs bosons as well as excited quarks and leptons are 
all predicted to be of the order of the composite mass scale, say 1 TeV. It strongly suggests that the excess of di-boson 
resonance events recently found by the ATLAS Collaboration \cite{ATLASanomaly} at about 2 TeV may be explained by the productions 
and decays of the first excited state of either one of the weak and Higgs bosons, and the glueballs. In the end of 
the last century, the anomalous events which might have indicated the substructure of the weak bosons, or quarks and leptons 
were reported from LEP \cite{ExcitedLepton} and from Tevatron and HERA \cite{Substructure}, but they have all disappeared with the higher statistics.
If the new anomaly reported from LEP stays still in the future experiments, it would indicate the substructure of 
the fundamental particles, one of the greatest discoveries in science in the 21st century!

\vspace{3mm}
\noindent
{\it Note added}:
Very recently after submitting this paper for publication, the ATLAS and CMS Collaborations reported the di-photon excess at 750 GeV \cite{ATLASGammaGamma,CMSGammaGamma}, which would indicate another evidence for the substructure of di-boson resonances whose masses are of the order of 1 TeV!

\vspace{3mm}
\noindent
\centerline{\textbf Acknowledgements}
One of the authors (H.T.) thanks the late Professor Yoichiro Nambu, to whom this paper has been dedicated, for many advices which he has received since he first met the great leader of particle theory in the 20th century in 1971.


\end{document}